\documentclass[preprintnumbers,showpacs,amsmath,amssymb,floatfix,prd,onecolumn,superscriptaddress,nofootinbib]{revtex4}
\usepackage{graphicx}
\usepackage{epsfig}
\usepackage{bm}
\usepackage{amsfonts}
\usepackage{epstopdf}
\usepackage{bm}
\usepackage{amsthm}
\usepackage{multirow}
\usepackage{subfigure}
\usepackage{verbatim}

\begin{document}

\title{Solar System Tests of a New Class of $f(z)$ Theory}

\author{Ji-Yao Wang}
\email{wjykana@foxmail.com}
\affiliation{Division of Mathematical and Theoretical Physics, Shanghai Normal University, 100 Guilin Road, Shanghai 200234,  P.R.China}

\author{Chao-Jun Feng}
\email{fengcj@shnu.edu.cn}
\affiliation{Division of Mathematical and Theoretical Physics, Shanghai Normal University, 100 Guilin Road, Shanghai 200234,  P.R.China}

\author{Xiang-Hua Zhai}
\email{zhaixh@shnu.edu.cn}
\affiliation{Division of Mathematical and Theoretical Physics, Shanghai Normal University, 100 Guilin Road, Shanghai 200234,  P.R.China}

\author{Xin-Zhou Li}
\email{kychz@shnu.edu.cn}
\affiliation{Division of Mathematical and Theoretical Physics, Shanghai Normal University, 100 Guilin Road, Shanghai 200234, P.R.China}

\begin{abstract}
Recently, a new kind of $f(z)$ theory is proposed to provide a different perspective for the development of reliable alternative models of gravity in which the $f(R)$ Lagrangian terms are reformulated as polynomial parameterizations $f(z)$. In the previous study, the parameters in the $f(z)$ models have been constrained by using cosmological data. In this paper, these models will be tested by the observations in the solar system. After solving the Ricci scalar as a function of the redshift, one could obtain $f(R)$ that could be used to calculate the standard Parameterized-Post-Newtonian (PPN) parameters. First, we fit the parametric models with the latest cosmological observational data. Then  the tests are performed by solar system observations. And last we combine the constraints of solar system and cosmology together and reconstruct the $f(R)$ actions of the $f(z)$ parametric models.

\end{abstract}

%\pacs{14.80.bn, 98.80.Es, 98.80.Cq }
\maketitle

%##############################################

\maketitle

%###########################################################################%
%Introduction
%###########################################################################%

\section{Introduction}
Einstein's general relativity (GR) has been successful in predicting many  phenomenologies in the universe and the solar system. In the past 20 years, more and more  astronomical observations have strongly confirmed that  the universe is under accelerating expansion\cite{Riess:1998cb,Perlmutter:1998np,Spergel:2003cb,Eisenstein:2005su,Kowalski:2008ez,Aghanim:2018eyx,Hinshaw:2012aka}. However, ordinary matters can only drive a decelerating universe. To explain the accelerating, a kind of exotic component in the universe is needed called the dark energy.

Another way to drive the accelerating expansion of the universe is to modify Einstein's gravity theory.The $f(R)$ theory is a kind of such modified gravity theories, in which, the  Einstein-Hilbert action is replaced by a function of $f(R)$. When $f(R)=R$, it is just the Einstein' gravity. Such a modified theory gives a geometrical explanation for the accelerating expansion of the universe \cite{Sotiriou:2008rp,Sotiriou:2008ve,Clifton:2006kc,Dunsby:2010wg}. Some famous $f(R)$ models have been deeply studied, such as  $R+\alpha R^2$ \cite{Teyssandier:1983zz}, $R+\mu/R$\cite{Dick:2003dw}, see also \cite{Sawicki:2007tf}. Recently, a new kind of $f(R)$ theory is proposed\cite{Lazkoz:2018aqk}, in which the $f(R)$ Lagrangian terms are reformulated as a polynomial parameterizations $f(z)$. It provides a new and different perspective for the  development of reliable alternative models of gravity. Cosmological data have been used to constrain the parameters in the $f(z)$ parametric models.

There are many experiments that could be used to test  gravity theories in a relatively high accurate level, including those in the solar system\cite{Jin:2006if,Berry:2011pb,Lin:2016nvj}, such as the gravitational redshift\cite{Lebach:1995zz}, the perihelion advance of Mercury\cite{Gai:2012ws}, the Shapiro time delay\cite{Shapiro:2004zz} and the Nordevert Effect\cite{Nordtvedt:1968qr}. As is known to all, general relativity is well consistent with the solar system tests. The parametric post Newtonian\cite{Nordtvedt:1972zz} limit measures the deviations of modified theories of gravity with respect from the general relativity, and it connects the observations with  some parameters in the gravitational potential, i.e. the Parameterized-Post-Newtonian (PPN) parameters. Therefore, it has become a useful framework to test the theories of gravity in the solar system. In this paper, after solving the  Ricci scalar  as a function of the redshift, one could obtain $f(R)$ and then to calculate the standard PPN parameters. 

Thus, in order to put constraints on the parameters of the models, we may take three steps:

1. Cosmology constraints;

2. Solar system tests;

3. Combining the data of solar system and cosmology together,  which is actually an
optimization problem.

The structure of this paper is as follows. In Section \ref{sec:fzr}, we obtain the equation for the Ricci scalar as a function of the redshift. And then, we solve this equation for each model proposed in Ref.\cite{Lazkoz:2018aqk}, in which every parametric model has an explicit formalism of $f(z)$. In Section \ref{sec:cos}, we fit the parametric models with latest cosmological observation. In Section \ref{sec:test}, we perform solar system constraints to these parametric models. The influence of the variations of parameters in the models is also discussed.  Next, in Section \ref{sec:comb} we combine the constraints of solar system and cosmology. We also reconstruct the action of the parametric models in Section \ref{sec:rec}. Finally, discussions and conclusions will be given in Section \ref{sec:conclusion}.

\section{From $f(z)$ parametric models to $f(R)$}\label{sec:fzr}

The most general $f(R)$ modified gravity theory is described by the following action:
\begin{equation}
\mathcal{S}=\int d^4x \sqrt{-g}\bigg[f(R)+\mathcal{L}_m\bigg]\,,
\end{equation}
where $g$ is the metric determinant and $\mathcal{L}_m$ is the Lagrangian of matter component. Here we use the units $8\pi G=1$. By varying the action with respect to the metric $g_{\mu\nu}$, one obtains the equations of motion as
\begin{equation}\label{eq:eom}
R_{\mu\nu}f_R-\frac{1}{2}g_{\mu\nu}f+(g_{\mu\nu}\nabla_\alpha\nabla^{\alpha}-\nabla_\mu\nabla_\nu)f_R=T_{\mu\nu}^m\,,
\end{equation}
where $f_R\equiv df/dR$ and $T^m_{\mu\nu}$ is the stress energy tensor of the matter. The FRW metric that describes a homogeneous and isotropic flat universe is given by
\begin{equation}
ds^2 = -dt^2 + a(t)^2\bigg[dr^2+r^2(d\theta^2 + \sin^2\theta d\phi^2)\bigg]\,,
\end{equation}
where $a(t)$ is the scale factor. From Eq.(\ref{eq:eom}) with the FRW background, one can obtain the modified Friedmann equations:
\begin{eqnarray}
H^2&=&\frac{1}{3f_R}\bigg(\rho_m+\frac{Rf_R-f}{2}-3H\dot{R}f_{2R}\bigg)\,,\label{eq:eom1}\\
	-3H^2-2\dot{H}&=&\frac{1}{f_R}\bigg[\dot{R}^2f_{3R}+(2H\dot{R}+\ddot{R})f_{2R}+\frac{1}{2}(f-Rf_R)\bigg]\,,\label{eq:eom2}
\end{eqnarray}
where $\dot=\frac{d}{dt}$. And the equations can be also read in terms of redshift:
\begin{eqnarray}
	H^2&=&\frac{R_z}{3f_z}\bigg[\frac{1}{2}\bigg(\frac{Rf_z}{R_z}-f\bigg)+\frac{3(1+z)H^2(R_zf_{2z}-R_{2z}f_z)}{R_z^2}\bigg]+\frac{R_z\Omega_m(1+z)^3}{f_z}\,.
\end{eqnarray}
In Ref.\cite{Lazkoz:2018aqk}, the authors have expressed $f(R)$ as a expressions of the redshift $z$, with $1+z=1/a$. The derivatives of $f(R)$ with respect to $R$, and of $R$ with respect to time are provided in terms of derivatives with respect to the redshift $z$. Therefore, one can obtain the Hubble parameter $H(z)$ from a given $f(z)$ parametric model, and use cosmological observational data such as the Type Ia Supernovae to constrain the parameters in these $f(z)$ models. In Ref.\cite{Lazkoz:2018aqk}, the authors have suggested eight ansatzes for $f(z)$ that are well-motivate:
\begin{eqnarray}
f(z)_{\textbf{Model} 1}&=&f_0 + f_3 (1+z)^3 \,,\label{eq:m1}\\
f(z)_{\textbf{Model} 2}&=&f_0 + f_1(1+z)+f_2 (1+z)^{2} +f_3(1+z)^3\,,\label{eq:m2}\\
f(z)_{\textbf{Model} 3}&=&f_0 + f_2(1+z)^2+f_3(1+z)^3\,,\label{eq:m3}\\
f(z)_{\textbf{Model} 4}&=&f_0 + f_1(1+z) + f_3(1+z)^3\,,\label{eq:m4}\\
f(z)_{\textbf{Model} 5}&=&f_{12} (1+z) ^ {1/2}+f_3(1+z)^3\,,\label{eq:m5}\\
f(z)_{\textbf{Model} 6}&=&f_{12} (1+z) ^ {1/4}+f_1(1+z)+f_2(1+z)^2+f_3(1+z)^3\,,\label{eq:m6}\\
f(z)_{\textbf{Model} 7}&=&f_{14} (1+z) ^ {1/4}+f_3(1+z)^3\,,\label{eq:m7}\\
f(z)_{\textbf{Model} 8}&=&f_{14} (1+z) ^ {1/4}+f_1(1+z)+f_2(1+z)^2+f_3(1+z)^3\,,\label{eq:m8}
\end{eqnarray}
where $f_{i}, i\in\{0,1,2,3,12,14\}$ are constant coefficients determined by observations. In the following, we call the above as Model $1\sim8$.

In this paper, we would like to consider $f(z)$ as parametric models. In other words, the models are still $f(R)$ models, though it is not easy to obtain the expressions in $f(R)$ theories. Thus, we can apply them in spherically symmetric geometry to test these $f(z)$ parametric models in the solar system observations. So the function of $f(R)$ should be solved for a given $f(z)$ parametric models. Then, by using the Eqs.(\ref{eq:eom1}) and (\ref{eq:eom2}), we eliminate the Hubble parameter $H(z)$ and get the equation of $R$ as the following:
\begin{eqnarray}
&&D_0 (R_{3z}R_z^2-2 R_{2z}^2R_z)
+ R_{3z}R_zR
-3 R_{2z}^2R+ D_3  R_{2z} R_z^2+D_5 R_{2z}R_zR + D_6R_z^3 +D_7R_z^2R = 0\,,\label{eq:eqr}
\end{eqnarray}
where
\begin{eqnarray*}
	D_0&=& \frac{(2\rho_m-f)}{f_z}\,,\\
	D_3&=&
	\frac{4 f_{2 z} \rho _m}{f_z^2}+\frac{2 \rho _m}{f_z(1+z)}-\frac{4 f}{(1+z) f_z}-\frac{2 f f_{2 z}}{f_z^2}\,,\\
	D_5&=& \frac{4 f_{2 z}}{f_z}+\frac{1}{1+z}\,,\\
	D_6&=&\frac{2 \rho _m}{(1+z)^2 f_z}-\frac{2 f_{3 z} \rho _m}{f_z^2}-\frac{2 f_{2 z} \rho _m}{(1+z)f_z^2}+\frac{4 f f_{2 z}}{(1+z) f_z^2}+\frac{f f_{3 z}}{f_z^2}-\frac{4 f}{(1+z)^2 f_z}\,,\\
	D_7&=&-\frac{f_{2 z}^2}{f_z^2}-\frac{f_{2 z}}{f_z(1+z)}-\frac{f_{3 z}}{f_z}+\frac{2}{(1+z)^2}\,.
\end{eqnarray*}
The subscript $z$ denotes the derivatives with respect to the redshift $z$, i.e. $f_{z}=df/dz, f_{2z}=d^2f/dz^2, f_{3z}=d^3f/dz^3$ and $R_{z}=dR/dz, R_{2z}=d^2R/dz^2, R_{3z}=d^3R/dz^3$. Therefore, for a given $f(z)$ model, one obtains $R(z)$ from Eq.(\ref{eq:eqr}), then  equations $(f,R)=(f(z), R(z))$ form a parametric representation of the function $f(R)$. Here we have used the following relations:
\begin{eqnarray}
	R
	&=&-3(H^2)_z(1+z)+12H^2\,,\label{eq:R0}\\
	R_z&=&9(H^2)_z-3(1+z)(H^2)_{2z}\,,\\\
	R_{2z}&=&6(H^2)_{2z}-3(1+z)(H^2)_{3z}\,.
\end{eqnarray}
and
\begin{eqnarray}
f_R &=& R_z^{-1}f_z\,,\\ \label{eq:fr}
f_{2R}& =& \left(f_{2z}R_z-f_zR_{2z} \right) R _z^{-3}\,,\\\label{eq:f2r}
f_{3R}& =& \frac {f_{3z}} {R_z^3} - \frac {f_z R_{3z} + 3f_{2z} R_{2z} } {R_z^4} + \frac {3f_z R_{2z}^ 2} {R_z^5}\,.
\end{eqnarray}
The Friedmann equation becomes:
\begin{eqnarray}
H^2 = \frac{D_0R_z+R}{6} \bigg[  1- (1+z) \left(\frac{f_{2z}}{f_z}-\frac{R_{2z}}{R _z} \right) \bigg]^{-1} \,.
\end{eqnarray}

\section{Observational Constraints from Latest Cosmological Observation}\label{sec:cos}

In this section, we fit the models by using the "joint light-curve analysis" (JLA)  sample, which contains $740$  spectroscopically confirmed type Ia supernovae with high quality light curves. We also use the cosmic microwave background (CMB) data and the baryon acoustic oscillations (BAO) data\cite{Feng:2015awr,Feng:2012gr}.

\subsection{The Data Fitting of Latest Cosmological Observation}

\subsubsection{SNeIa Constraints}

In this analysis, the distance estimator assumes hat supernovae with identical color, shape and galactic environment have on average the same intrinsic luminosity for all redshifts. This hypothesis is quantified by a linear model, yielding a standardized distance modulus\cite{Betoule:2014frx,Shafer:2015kda,Cheng:2018nhz}
\begin{equation}\label{equ:modulusobs}
\mu_{\text{obs}} = m_{\text{B}} - (M_{\text{B}} - \alpha \cdot s + \beta \cdot C + P \cdot \Delta_M)  \,,
\end{equation}
where $m_{\text{B}}$ is the observed peak magnitude in rest-frame B band, $M_{\text{B}}, s, C$ are the absolute magnitude, stretch and color measures, which are specific to the light-curve fitter employed, and $P(M_* >10^{10} M_\odot)$ is the probability that the supernova occurred in a high-stellar-mass host galaxy. The stretch, color, and host-mass coefficients ($\alpha, \beta, \Delta_M$, respectively) are nuisance parameters that should be constrained along with other cosmological parameters. On the other hand, the distance modulus predicted from a cosmological model  for a supernova at redshift $z$ is given by
\begin{equation}\label{equ:modulusmod}
\mu_{\text{model}} (z, \vec\theta)= 5\log_{10} \left[ \frac{D_L(z)}{10\text{pc}}\right]\,,
\end{equation}
where $\vec\theta$ are the cosmological parameters in the model, and $D_L(z)$ is the luminosity distance
\begin{equation}\label{equ:lum}
D_L(z) = (1+z)  \frac{c}{H_0} \int_0^z \frac{dz'}{E(z')} = (1+z)r_A(z)\,,
\end{equation}
for a flat FRW Universe. Here $r_A(z)$ is the comoving angular diameter distance. The $\chi^2$ statistic is then calculated in the usual way
\begin{equation}\label{equ:chi2sn}
\chi^2_{\text{SN} } = (\vec \mu_{\text{obs}}  - \vec \mu_{\text{model}})^T \mathbf{C_{\text{SN}}}^{-1}  (\vec \mu_{\text{obs}}  - \vec \mu_{\text{model}})\,,
\end{equation}
with $\mathbf{C_{\text{SN}}}$ the covariance matrix of $\vec \mu_{\text{obs}}$.

\subsubsection{CMB Data}

The CMB temperature power spectrum is  sensitive to the matter density, and it also measures precisely the angular diameter distance at the last-scattering surface, which is defined as
\begin{equation}\label{equ:angu}
\theta_{*} = \frac{r_s(z_*)}{r_A(z_*)} \,,
\end{equation}
where $r_s(z)$ is the comoving sound horizon
\begin{equation}
r_s(z_*) = \frac{c}{H_0}\int_0^{a(z_*)} \frac{c_s(a)}{a^2 E(a)} da \,,
\end{equation}
with the sound speed $c_s(a)$ given by
\begin{equation}
c_s(a) = \bigg[ 3 \left( 1+ \frac{3\Omega_{b0}h^2}{4\Omega_{\gamma 0}h^2} a \right) \bigg]^{-1/2} \,.
\end{equation}
Usually, $\theta_*$ is approximated based on the fitting function of $z_*$ given in Ref.~\cite{Hu:1995en}:
\begin{equation}
z_* = 1048 \left[1+0.00124(\Omega_{b0}h^2)^{-0.738}\right]\left[ 1+ g_1(\Omega_{m0}h^2)^{g_2}\right] \,,
\end{equation}
where
\begin{equation}
g_1 = \frac{0.0783(\Omega_{b0}h^2)^{-0.238}}{1+39.5(\Omega_{b0}h^2)^{0.763}} \,, \quad
g_2 = \frac{0.560}{1+21.1(\Omega_{b0}h^2)^{1.81} } \,.
\end{equation}
In CosmoMC package, the approximated $\theta_*$ is denoted as $\theta_{\text{MC}}$. In this paper, we fix $\Omega_{\gamma0} = 2.469\times 10^{-5} h^{-2}$,  and then the total radiation energy density, namely the sum of photons and relativistic neutrinos is  given by $\Omega_{r0} = \Omega_{\gamma0}(1+0.2271 N_{\text{eff}}) $,
where $ N_{\text{eff}}$ is the effective number of neutrino species, and the current standard value is $ N_{\text{eff}}=3.046$.  In the following, we use the Planck measurement of the CMB temperature fluctuations and the WMAP measurement of the large-scale fluctuations of the CMB polarization. This CMB data are often denoted by "Planck + WP". The geometrical constraints inferred from this data set are the present value of baryon density $\Omega_{b0}h^2$ and dark matter $\Omega_{m0}h^2$, as well as $100\theta_{\text{MC}}$. Thus, the $\chi^2 $ of the CMB data is constructed as
\begin{equation}\label{equ:chi2cmb}
\chi^2_{\text{CMB}} = (\nu - \nu_{\text{CMB}})^T \mathbf{C}^{-1}_{\text{CMB}} (\nu - \nu_{\text{CMB}})\,,
\end{equation}
where $\nu_{\text{CMB}}^T = (\Omega_{b0}h^2, \Omega_{m0}h^2, 100\theta_{\text{MC}}) ^T= (0.022065, 0.1199, 1.04131)^T$, and the best fit covariance matrix for $\nu$ is given by\cite{Betoule:2014frx,Ade:2013zuv}
\begin{eqnarray}
\mathbf{C}_{\text{CMB}} & = & 10^{-7}\left(\begin{array}{ccc}
0.79039 & -4.0042 & 0.80608\\
-4.0042 & 66.950 & -6.9243\\
0.80608 & -6.9243 & 3.9712\end{array}\right)\, .
\end{eqnarray}
after marginalized over all other parameters.

\subsubsection{BAO Data}
The BAO measurement provides a standard ruler to probe the angular diameter distance versus redshift by performing a spherical average of their scale measurement, which contains the angular scale and the redshift separation: $d_z = r_s(z_d)/D_V(z) $, where $r_s(z_d)$ is the comoving sound horizon at the baryon drag epoch,  and $D_V(z)$ is given by
\begin{equation}
D_V(z) \equiv \bigg[ r_A^2(z)\frac{ c z }{H(z)}\bigg]^{1/3} \,.
\end{equation}
The redshift of the drag epoch can be approximated by the following fitting formula,
\begin{equation}
z_d = \frac{ 1291(\Omega_{m0}h^2)^{0.251} }{ 1+ 0.659(\Omega_{m0}h^2)^{0.828} } \bigg[ 1+ b_1(\Omega_{b0}h^2)^{b_2}\bigg] \,,
\end{equation}
with
\begin{eqnarray}
b_1 &=& 0.313(\Omega_{m0}h^2)^{-0.419} \left[1 + 0.607(\Omega_{m0}h^2)^{0.674}\right], \nonumber\\
b_2 &=& 0.238(\Omega_{m0}h^2)^{0.223}  \,.
\end{eqnarray}
see, Ref. \cite{Eisenstein:1997ik}.  In the following, we will use the measurement of the BAO scale from Ref.~\cite{Beutler:2011hx,Padmanabhan:2012hf,Anderson:2012sa} and then the $\chi^2 $ of the BAO data is constructed as
\begin{equation}\label{equ:chi2bao}
\chi^2_{\text{BAO}} = (d_z - d_z^{\text{BAO}})^T\mathbf{C}^{-1}_{\text{BAO}} (d_z - d_z^{\text{BAO}}) \,,
\end{equation}
with $d_z^{\text{BAO}}=(d_{0.106},d_{0.35},d_{0.57})^T=(0.336,0.1126,0.07315)^T$, and the covariance matrix, also see Ref.~\cite{Betoule:2014frx}
\begin{eqnarray}
C_{\text{BAO}}^{-1} & = & \left(\begin{array}{ccc}
4444 & 0 &0 \\
0 & 215156 &0 \\
0&0& 721487\end{array}\right)\,.
\end{eqnarray}

\subsection{Fitting Results}

The  best-fit parameters by using the data of SNeIa+CMB+BAO is presented in Table \ref{table:hist1} and Table \ref{table:hist2},  in which we have also shown the best fit parameter values with 1$\sigma$ errors and the corresponding values of $\chi^2_{\text{min}}$.  Here we have also presented the fitting results of LCDM(Model 1) in Table \ref{table:hist1} as comparation \cite{Lazkoz:2019ivd,Benetti:2019gmo}.

\begin{table}[h]
	\centering
	\begin{tabular}{c|c|c|c|c}
	\hline
	\hline
	\multirow{2}{*}{  Parameters} & \multicolumn{4}{c}{ Parametric Models  } \\
	\cline{2-5}
    &  LCDM (Model 1)	& Model 2 &  Model 3&  Model 4    \\
	\hline
	\hline
	$\Omega_{m0}$&$0.257\pm 0.009$
	& $0.255\pm 0.010$  	& $0.257\pm0.010$ &  $0.256\pm 0.009$\\
	\hline
	$f_0$& $4.46\pm 0.09$
	& $4.46\pm0.12 $  	& $4.45\pm0.10$ &  $4.47\pm 0.09$ \\
	$f_1(10^{-2})$& $-$
	& $-0.34\pm2.71$  	& $-$ &  $-0.36\pm0.77 $ \\
	$f_2(10^{-2})$& $-$
	& $-0.66\pm0.74 $  	& $-0.66\pm2.10$ &  $-$ \\
	$f_{12}$& $-$
	& $-$ 	& $-$ &  $-$ \\
	$f_{14}$& $-$
	& $-$  	&$-$ &  $-$ \\
	
	$f_3$& $0.76\pm 0.03$
	& $0.76\pm 0.03$  	& $0.77\pm0.04$ &  $0.77\pm 0.03$ \\
	\hline
	$H_0$&$68.03\pm 0.73$
	& $68.53\pm 1.21$ & $68.55\pm1.24$ & $68.55\pm 1.28$ \\
	\hline
	$\alpha$&$0.141\pm 0.006$
	& $0.141\pm 0.006$ & $0.141 \pm 0.006$& $0.141\pm 0.006$ \\
	$\beta$&$3.09\pm 0.074$
	& $3.102\pm 0.073$ & $3.103\pm0.075$ & $3.102\pm 0.074$ \\
	$M_B$&$-19.11\pm 0.025$
	& $-19.10\pm 0.035$ & $-19.09\pm0.023$  &$-19.09\pm 0.026$ \\
	$\Delta M$&$-0.07\pm 0.023$
	& $-0.070\pm 0.023$ & $-0.069\pm0.024$ & $-0.070\pm 0.023$ \\
	\hline
	\hline
	$\chi^2_{min}/d.o.f$&$684.081/739$
	& $683.813/737$ & $683.815/738$  & $683.836/738$\\
	\hline
	\hline
\end{tabular}
	\caption{\label{table:hist1} Best fitting parameters for LCDM(Model 1) and Model 2-4.}
\end{table}

\begin{table}[h]
	\centering
	\begin{tabular}{c|c|c|c|c}
		\hline
		\hline
		\multirow{2}{*}{  Parameters} & \multicolumn{4}{c}{ Parametric Models  } \\
		\cline{2-5}
		 & Model 5 &  Model 6 &  Model 7 &  Model 8   \\
		\hline
		\hline
		$\Omega_{m0}$
		&$0.275\pm 0.002$&$0.257\pm 0.009$&$0.255\pm 0.003$&$0.265\pm 0.009$ \\
		\hline
		$f_0$
		&$-$&$-$&$-$&$-$ \\
		$f_1(10^{-2})$
		&$-$&$0.09\pm 0.56$&$-$&$0.028\pm 0.035$ \\
		$f_2(10^{-2})$
		&$-$&$0.008\pm0.090 $&$-$&$-0.018\pm 0.08$ \\
		$f_{12}$
		&$1.83\pm 0.19$&$1.55\pm 0.23$&$-$&$-$ \\
		$f_{14}$
		&$-$&$-$&$2.06\pm 0.24$&$2.08\pm 0.31$ \\
		
		$f_3$
		&$0.83\pm 0.01$&$0.77\pm 0.02$&$0.80\pm 0.03$&$0.76\pm 0.01$ \\
		\hline
		$H_0$
		&$64.56\pm 0.63$&$68.54\pm 1.28$&$66.35\pm 0.68$&$68.54\pm1.28 $\\
		\hline
		$\alpha$
		&$0.140\pm 0.006$&$0.141\pm 0.006$&$0.141\pm 0.006$&$0.140\pm 0.006$\\
		$\beta$
		&$3.081\pm 0.074$&$3.102\pm 0.075$&$3.102\pm 0.074$&$3.089\pm 0.074$ \\
		$M_B$
		&$-19.18\pm 0.024$&$-19.09\pm 0.035$&$-19.09\pm 0.041$&$-19.14\pm 0.034$\\
		$\Delta M$
		&$-0.072\pm 0.023$&$-0.070\pm 0.023$&$-0.070\pm 0.002$&$-0.0.071\pm 0.027$\\
		\hline
		\hline
		$\chi^2_{min}/d.o.f$
		&$697.341/739$&$683.621/737$&$687.925/739$&$683.718/737$\\
		\hline
		\hline
	\end{tabular}
	\caption{\label{table:hist2} Best fitting parameters for Model 5-8.}
\end{table}

 From the fitting results one can see that  the differences of $\chi^2_{\text{min}}$ among model 2, 3, 4, 6, 8 and LCDM not so obvious.

\section{Observational Constraints from the Solar System}\label{sec:test}
In this section, starting from the definitions of parametrized Post-Newtonian(PPN) formalism, we will first give a brief review on the PPN formalism of $f(R)$ theories, and then we will give an example to show how to calculate the PPN parameters. Finally, we will performance some observational tests on Model 2-8 from Eqs.(\ref{eq:m1})$-$(\ref{eq:m8}) by parametrized Post-Newtonian Formalism of $f(z)$ Parametric models in terms of analytic fuction $f(z)$ and its derivatives.

\subsection{Brief Review on the Parametrized Post-Newtonian Formalism of $f(R)$ Theory}

The GR theory is very successful in predicting the behavior of the gravitational phenomena in the solar system, so every kind of generation of GR proposed in order to explain the accelerating expansion of the universe, such as the $f(R)$ theories, should be tested in the solar system. 

With the improvement of observation technology, it is important to distinguish one metric theory from another that may lead to different observable effects. The Post-Newtonian limit provides a simple way to compare different metric theories when one takes the slow-motion, weak-field limit and the Parametrized Post-Newtonian (PPN) formalism has become a basic tool to connect alternative metric gravitational theories with solar system experiments.

Under this assumption, one usually expands about the GR solutions up to some perturbation orders when taking into the account deviation from GR. In an environment of high density such as the Sun, the Ricci scalar is larger than the cosmological background. In the following, we take the standard PPN\cite{Nordtvedt:1972zz} expansion of the Schwarzschild metric:
\begin{eqnarray}
ds^2=-\left[1-2\frac{GM}{r} + 2(\beta-\gamma) \left(\frac{GM}{r}\right) ^2 \right] dt^2+\left[1+2 \gamma \frac{GM}{r} \right]dr^2+ r^2 d\Omega^2 \,,
\end{eqnarray}
where $\alpha$, $\beta$ and $\gamma$ are dimensionless parameters known as the Edditon parameters, which describe the deviations from GR. It is evident that the standard GR solution corresponds to the case $\beta=\gamma=1$. The parameter $\gamma$ measures how the space is curved by unit mass and it is also connected with time delay or the effect of light deflection, while the parameter $\beta$ measures how much the non-linearity is in gravitational superposition, which can be measured though Nordtvedt effect and the perihelion shift.

The PPN parameters defined in scalar-tensor theories can be obatained by the ordinary method to build the PPN formalism\cite{Damour:1992we}. And scalar-tensor theories turns out to be equivalent to $f(R)$ theories by some replacements. Thus, analogy between scalar-tensor and higher-order gravity, one can obtain the PPN formalism\cite{Capozziello:2006jj,DeLaurentis:2009yd}, 
\begin{eqnarray}\label{gb1}
\gamma-1&=&-\frac{f''(R)^2}{f'(R)+2f''(R)^2}\,,\label{gb11}\\
\beta-1&=&\frac{1}{4}\left( \frac{f'(R)\cdot f''(R)}{2f'(R)+3f''(R)^2} \right)\frac{d\gamma}{dR}\,,\label{gb12}
\end{eqnarray}
which provides  a simple and easy way to constraint the different kinds of $f(R)$ theories from solar system experiments when the analytic expression of $f(R)$ is known.  Usually, the function of $f(R)$ could be only expressed as a parametric function, e.g. that solved form Eq.(\ref{eq:eqr}),  then we reformulate the expression of $\gamma$ and $\beta$ as the following: 
\begin{eqnarray}\label{eq:gamma&beta}
\gamma - 1& = & -\frac{\xi(\lambda)^2 }{f_\lambda R_\lambda^5+2\xi(\lambda)^2 } \,,\label{eq:gamma}\\
\beta - 1 & =& \frac{1}{4}\bigg[ \frac{f_\lambda R_\lambda\xi(\lambda)}{2f_\lambda R_\lambda^5+3\xi(\lambda)^2}\bigg]\gamma_\lambda \,, \label{eq:gamma1}\\
\gamma_\lambda&\equiv& \frac{d \gamma}{d \lambda} = -\frac{2\xi(\lambda)\xi(\lambda)_\lambda }{f_\lambda R_\lambda^5+2\xi(\lambda)^2 }+ \frac{R_\lambda^4\xi(\lambda)^2(\xi(\lambda)+ 6f_\lambda R_{2\lambda})+4\xi(\lambda)^3\xi(\lambda)_\lambda}{(f_\lambda R_\lambda^5+2\xi(\lambda)^2)^2}\,,\label{eq:gamma3}
\end{eqnarray}
where we have introduced a parameter $\lambda$, and both $f(\lambda)$ and $ R(\lambda)$ are functions of $\lambda$, then $f(R)$ is represented as a parametric function. Here the  $\xi(\lambda)$ is defined by
\begin{equation}
\xi(\lambda) \equiv f_{2\lambda} R_\lambda - f _ {\lambda} R _{2\lambda}\,,
\end{equation}
and  we also have $\xi_\lambda = f_{3 \lambda} R_\lambda-f_\lambda R_{3\lambda}$. 

For example,  one can rewrite the model $f(R)=R+\alpha R^2$ as a parametric function 
\begin{eqnarray}
f&=& g(\lambda) \,,\\
R&=&\frac{\sqrt{1+4\alpha g(\lambda)}-1}{2\alpha}\,,
\end{eqnarray}
where $g(\lambda)$ is some function of $\lambda$. Then the1 PPN parameters could be obtained by using Eqs.(\ref{eq:gamma})-(\ref{eq:gamma3}):
\begin{eqnarray}
\gamma-1&=&-\frac{4\alpha^2}{1+8\alpha^2+2\alpha R }\label{gamma222}\,,\\
\beta-1&=&\frac{2\alpha^4(1+2\alpha R)}{(1+6\alpha^2+2\alpha R)(1+8\alpha^2+2\alpha R)^2}\label{beta222}\,,
\end{eqnarray}
which is just the same as the results from Eqs.(\ref{gb11}) and (\ref{gb12}), which do not depend on the parameter $\lambda$.

\subsection{Parametrized Post-Newtonian Formalism of $f(z)$ Parametric Models}

Among the $f(z)$ parametric models, Model $1$ has an exact solution with $f_0=6(1-\Omega_m), f_3=3\Omega_m$:
\begin{eqnarray}\label{eq:lmr}
R(z)=12(1-\Omega_m)+3\Omega_m(1+z)^3 \,,
\end{eqnarray}
where  $\Omega_i$ are the relative densities of the components and hereafter the subscript $m$ denotes the dust matter. So the function of $f(R)$ is
\begin{equation}\label{eq:lm}
f(R) = R -6(1-\Omega_m) \,,
\end{equation}
which is just the $\Lambda$CDM Model.

Since the values of $\gamma$ and $\beta$ do not depend on how to choice the parameter $\lambda$, we take $\lambda=z$ in the following, then $f(R)$ is a parametric function in terms of $z$:
\begin{eqnarray}
f&=& f(z)\\
R&=& R(z) \,.
\end{eqnarray}
By using the Eqs.(\ref{eq:m1}) and (\ref{eq:lmr}), we get 
\begin{equation}
\xi(z) = 0\,, \gamma=1\,, \beta = 1\,,
\end{equation} 
which are exactly the same as the results that calculated by substituting Eq.(\ref{eq:lm}) into Eqs.(\ref{gb11}) and (\ref{gb12}).

For Model $2-8$, one usually can not obtain the exact solution of $R(z)$ through Eq.(\ref{eq:eqr}), then the numerical approach is needed to solve this equation. To numerically solve Eq.(\ref{eq:eqr}), we take the same initial conditions as those in Ref.\cite{Lazkoz:2018aqk}. Once the solution of $R(z)$ is found, one can obtain $f(R)$ immediately. In fact, Model 2, 3 and 4 have no significant difference with Model 1. And as
it is reasonable, the  deviation of other models from Model 1 becomes remarkable when scalar curvature is large.

The uncertainties of the parameter $\gamma$ and $\beta$ are given by
\begin{eqnarray}\label{eq:uncer}
\sigma_\gamma = \sqrt{\sum_{i} \left(\frac{\delta \gamma}{\delta f_i}\right)^2\sigma^2_{f_i} } \,,\quad \sigma_\beta = \sqrt{\sum_{i} \left(\frac{\delta \beta }{\delta f_i}\right)^2\sigma^2_{f_i} } \,,\quad i \in \{0,1,2,3,12,14\} \,.
\end{eqnarray}
The variations of $\gamma, \beta$ can be obtained by using the following equations:
\begin{eqnarray}
\delta\gamma &=&\frac{R_z^4\xi}{(f_zR_z^5+2\xi^2)^2}\bigg[-2 f_zR_z^2\delta f_{2z} + R_z(2 f_z R _{2z}+\xi )\delta f_{z} - f_z(2 f_{2z}R_z- 5\xi)\delta R_z  +2 f_z^2R_z \delta R_{2z}\bigg] \,,\label{eq:v1} \\
\delta \beta & =&
\nonumber \frac{\gamma_z}{4(2f_zR_z^5+3\xi^2)^2}\bigg[ f_zR_z(2f_zR_z^5-3\xi^2)(R_z\delta f_{2z} - f_z \delta R_{2z} ) + R_z^2(3\xi^2f_{2z}-2f_z^2R_z^4R _{2z})\delta f_z \\
&&f_z^2(2R_z^6f_{2z}-3\xi^2R_{2z}-8R_z^5\xi)\delta R_z \bigg]+  \frac{1}{4}\bigg( \frac{f_zR_z\xi}{2f_zR_z^5+3\xi^2}\bigg)\delta\gamma_z \,, \label{eq:v2} \\
\delta\gamma_z\nonumber
&=&\frac{1}{{(f_zR_z^5+2\xi^2)^2}}\bigg[\bigg( 2\xi_z(f _zR_z^5+2\xi^2)+8\xi^2\xi_z-8\xi\frac{R_z^4\xi^2(\xi+6f_zR_{2z})+4\xi^3\xi_z}{f_zR_z^5+2\xi^2}\\
\nonumber
&&+(R_z^4\xi^2+2R_z^4\xi(\xi+6f_zR_{2z})+12\xi^2\xi_z)  \bigg)\delta \xi +\bigg( 4\xi^3+2\xi  \bigg)\delta \xi_z\\
\nonumber
&&+\bigg( 2\xi\xi_z R_z^5+6R_z^4 R_{2z}\xi^2-2\frac{R_z^4\xi^2(\xi+6f_zR_{2z})+4\xi^3\xi_z}{(f_zR_z^5+2\xi^2)}(R_z^5+4\xi)  \bigg)\delta f_z\\
&&+\bigg(  10\xi\xi_z f_zR_z^4-10\frac{R_z^4\xi^2(\xi+6f_zR_{2z}+4\xi^3\xi_z)}{(f_zR_z^5+2\xi^2)}f_zR_z^4+4R_z^3\xi^2(\xi+6f_zR_{2z}) \bigg)\delta R_{z}+6f_z\xi^2R_z^4 \delta R_{2z}\bigg]\,, \label{eq:v3}
\end{eqnarray}
where the variations of $f,f_z,f_{2z}$ and $f_{3z}$ could be easily obtained by using Eqs.(\ref{eq:m2}-\ref{eq:m8}). For instance,
\begin{eqnarray*}
	\delta f&=& \delta f_0 + \delta f_1(1+z) + \delta f_3(1+z)^3 \,,\\
	\delta f_z&=& \delta f_1+ 3\delta f_3(1+z)^2\,,\\
	\delta f_{2z}&=&  6\delta f_3(1+z)\,,\\
	\delta f_{3z}&=& 6 \delta f_3\,,
\end{eqnarray*}
for Model 2. However, to get the variations of  $R,R_z,R_{2z}$ and $R_{3z}$, one needs to solve the following equation for $\delta R$:
\begin{equation}\label{eq:pert}
A_0+A_1 \delta R_{3z} + A_2 \delta R_{2z} + A_3 \delta R_z +A_4 \delta R = 0\,,
\end{equation}
with the coefficients
\begin{eqnarray*}
	A_0&=&-\bigg[ (\hat{R}_{3z}\hat{R}_z^2-2 \hat{R}_{2z}^2\hat{R}_z)\delta D_0+   \hat{R}_{2z} \hat{R}_z^2 \delta D_3+ \hat{R}_{2z}\hat{R}_z\hat{R} \delta D_5 + \hat{R}_z^3 \delta D_6 + \hat{R}_z^2\hat{R} \delta D_7 \bigg]\,,\\
	A_1 &=& \hat{D}_0 \hat{R}_z^2+\hat{R}_z \hat{R}\,,\\
	A_2 &=& -4\hat{D}_0\hat{R}_z\hat{R}_{2z}+\hat{D}_3 \hat{R}_z^2+\hat{D}_5\hat{R}_z\hat{R}-6\hat{R}_{2z} \hat{R}\,,\\
	A_3 &=& -2\hat{D}_0 R_{2z}^2+2\hat{D}_3 \hat{R}_{2z}\hat{R}_z+\hat{D}_5\hat{R}_{2z}\hat{R}+3\hat{D}_6\hat{R}_z^2+2\hat{D}_7\hat{R}_z\hat{R}+\hat{R}_{3z}\hat{R}\,,\\
	A_4 &=& \hat{D}_5\hat{R}_{2z}\hat{R}_z+\hat{D}_7\hat{R}_z^2+ \hat{R}_{3z}\hat{R}_z-3\hat{R}_{2z}^2 \,,\\
\end{eqnarray*}
where
\begin{eqnarray}
\delta D_0&=&   \bigg(\hat{D}_0\delta f_z +\delta f\bigg)/\hat{f_z}\,,\\
\delta D_3&=& 2 \frac{\hat f_{2z}}{\hat f_z^2}\delta f+\frac{4\delta f}{\hat f_z(1+z)}+\frac{(4\rho_m-2\hat f)\hat f_{2z}}{\hat f_z^2}\left(\frac{\delta f_z}{\hat f_z}- \frac{\delta f_{2z}}{\hat f_{2z}} \right)+\hat{D_3}\frac{\delta f_z}{\hat f_z}\,,\\
\delta D_5&=&  \frac{4\hat f_{2z}}{\hat f_z^2}\delta f_z- \frac{4\delta f_{2z}}{\hat f_z}  \,,\\
\delta D_6&=&(2\rho_m-\hat f) \bigg(\frac{\delta f_{3z}}{\hat f_z^2} -\frac{\hat f_{3z}}{\hat f_z^3}\delta f_z \bigg)
+ \delta f\bigg(\frac{4}{\hat f_z(1+z)^2} -\frac{4\hat f_{2z}}{\hat f_z^2(1+z)} -\frac{\hat f_{3z}}{\hat f_z^2} \bigg)\\\nonumber
&&+(2\rho_m- 4\hat f)\bigg(\frac{\delta f_{2z}}{\hat f_z^2(1+z)}-\frac{\hat f_{2z}\delta f_z}{\hat f_z^3(1+z)} \bigg)+ \hat{D_6}\frac{\delta f_z}{\hat f_z}\,,\\
\delta D_7&=&  \frac{\delta f_{3z}}{\hat f_z} -\frac{\hat f_{3z}}{\hat f_z^2}\delta f_z +\frac{\delta f_{2z}}{\hat f_z(1+z)} -\frac{\hat f_{2z}}{\hat f_z^2(1+z)}\delta f_z +\frac{2\hat f_{2z}\delta f_{2z}}{\hat f_z^2}-2\frac{\hat f_{2z}^2}{\hat f_z^3}\delta f_z \,.
\end{eqnarray}
It is hardly to solve Eq.(\ref{eq:pert}) exactly, however, for Model~1, one could get the asymptotic solutions.  In the limit of $z\rightarrow 0$, the coefficients $A_0\sim A_4$ all become constants, so we have a constant solution
\begin{equation}
\delta R|_{z\rightarrow 0} = -\frac{A_0}{A_4} =\frac{ 4+15\Omega _m  }{ \Omega _m }\delta f_{3} +6(\delta f_0+\delta f_3)\,.
\end{equation}
In the limit of $z\rightarrow \infty$, $A_1$ is the most important coefficient, then Eq.(\ref{33}) becomes
\begin{eqnarray}\label{33}
\delta f_{3z} + \delta R_{3z}=0\,,
\end{eqnarray}
then we get
\begin{equation}
\delta R|_{z\rightarrow \infty} = -\delta f \,.
\end{equation}
Therefore, the asymptotic behavior of $\delta R$ is regular in Model~1. In fact, this conclusion is also valid in Model~2-8.

\subsection{Data Description of Solar System Experiments}

\subsubsection{VLBA Data}
To test the $f(z)$ parametric models Eqs.(\ref{eq:m2})-(\ref{eq:m8}) in the solar system, we use the data from the Very Long Baseline Array (VLBA) at 43, 23 and 15 GHz, in which the gravitational bending of radio waves is observed and then the Eddington parameter $\gamma-1$ is constrained by \cite{Fomalont:2009zg}:
\begin{equation}\label{eq:gamma2}
|\gamma-1| \leq  2\times10^{-4} \,.
\end{equation}
From the observations of the the perihelion advance of Mercury, $\beta-1$ is constrained by \cite{Will:2005va}:
\begin{equation}\label{eq:beta}
|\beta-1|\leq  0.0023 \,.
\end{equation}
\subsubsection{Nordtvert effect Data}
As is known, the Nordtvert effect\cite{Nordtvedt:1968qr}, as an effect that relates to the difference between the inertial mass $M(I)$ and the gravitational mass $M(G)$,
\begin{equation}
\frac{M(G)_i }{M(I)_i}=1-\eta_{\mathrm{N}} \frac{1}{M_i c^2} \int \frac{G\rho(\vec{r})\rho (\vec{r}' )\:d^3 rd^3 r'}{2\:|\vec{r}-\vec{r} '|}\,,
\end{equation}
can be described by  the combination of $\gamma$ and $\beta$\cite{Williams:2004qba}:
\begin{equation}\label{eq:nord}
\eta_{\mathrm{N}}= 4 \beta-\gamma-3\,,
\end{equation}
which could be observed by the Lunar Laser Ranging Tests (LLT). This parameter $\eta_{\mathrm{N}}$ could be regarded as  as another PPN parameter, which is constrained by \cite{Williams:2004qba}:
\begin{equation}\label{eq:eta}
-1.3 \times 10^{-4} \leq\eta_{\mathrm{N}}\leq 0.9\times 10^{-4} \,.
\end{equation}
We summarized these data in Table \ref{tab:data}.
\begin{table}[h]
	\begin{tabular}{ccccc}
		\hline\hline
		\textbf{PPN Parameters}  & \textbf{Related Phenomenon}          & \textbf{Experiment}    & \textbf{Result}&\textbf{Value in GR} \\ \hline\hline
		\multirow{2}{*}{$\gamma-1$} & Time Delay                           & Cassini mission        & $(2.1\pm2.3)\times10^{-5}$ \cite{Bertotti:2003rm}        &   \multirow{2}{*}{0}\\
		& Gravitational Bending of Radio Waves & VLBA                   & $\pm 2\times10^{-4}$\cite{Fomalont:2009zg}             \\ 
		$\beta-1$                  & Perihelion Advance of Mercury        & Solar System Ephemeris & $\pm 0.0023$\cite{Will:2005va}      &0         \\ 
		$\eta_{\mathrm{N}}$                      & Nordtvert Effect                     & LLT                    & $(-0.2\pm1.1)\times 10^{-4}$\cite{Hofmann:2018myc}     &0         \\ \hline\hline
	\end{tabular}
	\caption{The observational values of PPN parameters. \label{tab:data}}
\end{table}

Notice that in Table III, the boundary of PPN parameter  $\eta-1$ given by Cassini mission is contained in that of VLBA. So in the tests results, we may mark the PPN parameter $\eta-1$ of the models with a $\checkmark$ if they are favored by Cassini mission and $\bigcirc$ if they are disfavored by Cassini but favored by VLBA, while the $\triangle$  denotes  that the model is not in consistent with the solar system observations.

\subsection{Tests Results of Solar System}

We first calculate the PPN parameters by the value of $f_i$ given in ref.\cite{Lazkoz:2018aqk}. By taking the values of $f_i$ in Table~1 of Ref.\cite{Lazkoz:2018aqk},  one can obtain the values of PPN parameters with their uncertainty through Eqs.(\ref{eq:gamma})-(\ref{eq:v3}). We summarized the results in Table~\ref{tab:res}.
\begin{table}[h]
	\begin{tabular}{ccccccc}
		\hline\hline
		\textbf{}          & & \textbf{$\gamma-1$} & & \textbf{$\beta-1$}& &  $\eta_{\mathrm{N}}$ \\ \hline\hline
		\textbf{Model $2$} &   & $(-1.9271\pm0.0805)\times 10^{-6}  $   && $(2.7991\pm0.0778)\times 10^{-9} $& & $ (1.9330\pm0.0805 )\times 10^{-6}    $             \\ 
		\textbf{Model $3$} &   & $ (-1.7296\pm 0.0149)\times 10^{-6} $  & & $(1.280\pm0.2320)\times 10^{-10} $& & $( 1.7297\pm 0.0149  )\times 10^{-6} $              \\ 
		\textbf{Model $4$}  &  & $ (-1.5933\pm0.0057)\times 10^{-6}  $  & & $(8.454\pm1.7110)\times 10^{-10}  $ && $(1.5936 \pm 0.0057)\times 10^{-6}  $           \\ 
		\textbf{Model $5$}  &  & $ (-2.0565\pm0.0113)\times 10^{-5}   $  & & $(-6.4361\pm0.0144)\times 10^{-9}  $ && $( 2.0564\pm0.0113 )\times 10^{-5}  $              \\ 
		\textbf{Model $6$} &   & $ (-2.0471\pm0.0091)\times 10^{-5}  $  & & $(-6.3926\pm0.0148)\times 10^{-9}  $ && $ (2.0480\pm0.0091 )\times 10^{-5}    $           \\ 
		\textbf{Model $7$} &   & $ (-8.9343\pm 0.1201)\times 10^{-6} $ & & $(-1.7904\pm 0.0247)\times 10^{-9}  $ && $(8.9339 \pm 0.1201)\times 10^{-6}   $            \\ 
		\textbf{Model $8$}  &  & $ (-6.6759\pm0.1215 )\times 10^{-6} $ & & $(-1.2079\pm 0.0179)\times 10^{-9} $ && $( 6.6753 \pm0.1215)\times 10^{-6}   $                      \\ \hline\hline
	\end{tabular}
	\caption{The values of PPN parameters and their $1\sigma$ errors, where we use the values from Ref.\cite{Lazkoz:2018aqk}.\label{tab:res}} in the calculation.
\end{table}

In Table~\ref{tab:res}, the central values of  $\gamma-1, \beta-1 , \eta_{N}$ are obtained by using the best fitting values of $f_i$ in Ref.\cite{Lazkoz:2018aqk}. If the center value of one parameter, such as the $|\gamma-1|$, falls within the range given by Eq. (\ref{eq:gamma}) (\ref{eq:beta}) or (\ref{eq:eta}),  the corresponding model is regarded as being consistent with observations in respect of that parameter. From Table~\ref{tab:res2}, one can see that  Model 2-4  and Model 7-8 are favored by observations, while Model~5-6 are not consistent with the solar system observations and . And even when the $1\sigma$ uncertainty of these parameters are taken into account, Model~5-6 could still hardly favored by  observations in respect of $\eta_N$. We summarized the results in Table \ref{tab:res2}.  

\begin{table}[h]
	\begin{tabular}{ccccccc}
		\hline\hline
		\multicolumn{1}{l}{\multirow{2}{*}{}} & \multicolumn{2}{c}{\textbf{$\gamma-1$}}                 & \multicolumn{2}{c}{\textbf{$\beta-1$}}                 & \multicolumn{2}{c}{\textbf{$\eta_N$}}  \\ \cline{2-7}
		\multicolumn{1}{l}{}                  &$\,\,\,$ \textbf{Best Fit}$\,\,\,$ &$\,\,\,$ \textbf{Within $1\sigma$}$\,\,\,$& $\,\,\,$\textbf{Best Fit}$\,\,\,$ & \textbf{Within $1\sigma$} & $\,\,\,$\textbf{Best Fit}$\,\,\,$ &$\,\,\,$ \textbf{Within $1\sigma$}$\,\,\,$ \\ \hline\hline
		\textbf{Model$2$}    & $\checkmark$    &  $-$    & $\checkmark  $        & $-  $     & $\checkmark$         & $-$ \\ 
		\textbf{Model$3$}    &  $\checkmark$       & $-  $     & $\checkmark $      & $-  $      & $\checkmark $   & $- $            \\ 
		\textbf{Model$4$}   &  $\checkmark$       & $-  $        & $\checkmark  $     & $- $      & $\checkmark$ & $- $         \\ 
		\textbf{Model$5$}    & $\bigcirc$      & $\bigcirc$          & $\checkmark $   & $-  $     & $\checkmark $     & $-$    \\ 
		\textbf{Model$6$}   & $\bigcirc$    &  $\bigcirc$    & $\checkmark $     & $-  $    &$\checkmark $    & $-$        \\ 
		\textbf{Model$7$}    & $\bigcirc$        & $\bigcirc$    &  $\checkmark$&  $- $           & $\checkmark $     &  $- $              \\ 
		\textbf{Model$8$}    & $\bigcirc$      & $\bigcirc$    & $\checkmark $     &  $-  $    & $\checkmark $    & $- $       \\ \hline \hline

	\end{tabular}

	\caption{The test results of $f(z)$ parametric models. The $\checkmark$ denotes that the model is in consistent with the data of Cassini Mission, while the $\bigcirc$  denotes  that the model is not in consistent with Cassini Mission but is  favored by the data of  VLBA. The "$-$" sign means that the $1\sigma$ error is not necessarily considered while the center value is favored.  Here we use the values from Ref.\cite{Lazkoz:2018aqk} in the calculation.
		\label{tab:res2}}
\end{table}

From Eqs.(\ref{eq:m2})-(\ref{eq:m8}), one can see that all models have the parameter $f_3$. Therefore, we also check the changes of the PPN parameters with respect to $\delta f_3\sim 0.15$ in  for some models:
\begin{eqnarray}
\delta (\gamma-1)_{\text{Model 2}}&\sim& 0.4\times10^{-6}\,,\\
\delta (\gamma-1)_{\text{Model 5}}&\sim& 0.15\times10^{-5}\,,\\
\delta (\gamma-1)_{\text{Model 7}}&\sim& 0.3\times10^{-6}\,.
\end{eqnarray}
We found that the values of the PPN parameters changed a little while $|\delta f_3|$ is larger than its $1\sigma $ error. Therefore, the uncertainty of $\delta f_3$ can hardly change our results. We also checked other parameters in Model~2-8, and got the same conclusion. Since the PPN parameter $\beta-1$ of each model is  much less than the boundary given by experiments, we may only check how the perturbations work on $\gamma-1$.

Next we calculate the PPN parameters again by the best fitting values given in Table I and II form Section III. Since the PPN parameter $\beta-1$ is much less than the experimental boundary and is negligible in the constraints to the models according to above discussion, in order to achieve more general results, we may relax the condition to $\beta-1=0$. The test results are summarized in Table \ref{tab:res}.

\begin{table}[h]
	\begin{tabular}{ccccccc}
		\hline\hline
		\textbf{}           && \textbf{$\gamma-1$} &&Best Fit&&Within 1$\sigma$\\ \hline\hline
		\textbf{Model $2$}  &&    $(-2.2204\pm0.0728)\times 10^{-6} $ && $\bigcirc$     &&  $\bigcirc$     \\ 
		\textbf{Model $3$}  &&    $( -1.3128\pm 0.0480)\times 10^{-6}$  &&  $\checkmark$   && $- $      \\ 
		\textbf{Model $4$}  &&   $( -1.1759 \pm0.0663)\times 10^{-6}  $ &&  $\checkmark$    &&   $- $    \\ 
		\textbf{Model $5$}  &&   $ (-3.2257\pm0.0141)\times 10^{-5} $  &&    $\bigcirc$   &&$\bigcirc$    \\ 
		\textbf{Model $6$}  &&    $ (-3.3481 \pm0.0289)\times 10^{-5} $   &&   $\bigcirc$  && $\bigcirc$   \\ 
		\textbf{Model $7$}  &&    $ (-5.6740\pm 0.0271)\times 10^{-5}$ &&  $\bigcirc$  && $\bigcirc$   \\ 
		\textbf{Model $8$}  &&   $( -1.3028\pm0.0374) \times 10^{-5} $&&  $\bigcirc$   && $\bigcirc$     \\ \hline\hline
	\end{tabular}
	\caption{The values of PPN parameters and their $1\sigma$ errors,  where we use the values from  Table I in the calculation.\label{tab:ras}}
\end{table}

From the test results of the two groups of best fitting values we can see that Model 2, 3, 4, which are just minor modification of GR, are well favored with  solar system tests, while Model 5 and 6 does not perform so well in solar system test than that of Model 2-4. And though Model 7 and 8 do not  pass the test very well when we use the values from  Table I, they are still favored with the solar system test in the vision of VLBA. So in next Section, we will combine the constraints of solar system and cosmology together.

\section{Combining the Constraints of Solar System and Cosmology}\label{sec:comb}

In this section, we will combine the constraints of solar system and cosmology in the way of covex optimization:
\begin{eqnarray}
\text{min} &&\chi^2=\chi^2_{\text{SN Ia}}+\chi^2_{\text{CMB}}+\chi^2_{\text{BAO}}\,,\nonumber\\
\text{subject to}&&-2\times10^{-6}<\gamma-1<4.4\times10^{-5}\,,-0.0023<\beta-1<0.0023\,,-1.3\times10^{-4}<\eta_{\text{N}}<0.9\times10^{-4}\,,
\end{eqnarray}
where the value of $\gamma-1$ is given by the Cassini Mission. Considering the fitting results and solar system test above, we may choose Model 2:
\begin{eqnarray*}
f_{0}+f_1(1+z)+f_2(1+z)^2+f_3(1+z)^3
\end{eqnarray*}
as an  example. And we suppose that $f_3=\Omega_{m0}$ is always established, according to Eq.(\ref{eq:lmr}), Since that the values of PPN parameters dos not change obviously when $f_3$ changes according to Fig.3.

From Eq.(\ref{gb1}) and Eq.(\ref{eq:gamma}), one can obtain that
\begin{eqnarray}
(1-2\gamma)(\frac{R_{2z}}{R_z})^2-(1-2\gamma)(\frac{f_{2z}}{f_z})^2+2(2\gamma-1)(\frac{R_{2z}}{R_{z}})(\frac{f_{2z}}{f_z})-\frac{1}{f_{z}}(\gamma-1)R_z^3=0\,.
\end{eqnarray}
and one can see that the deviation of PPN parameter $\gamma-1$ is mainly contributed by the last term of the equation, which does not include the contribution of $f_0$ terms. So we should first consider how the value of PPN parameter may change with $f_1$ and $f_2$ respectively. We perform perturbation to $f_1$ and $f_2$ individually. From the perturbation view we find that the value of $\gamma-1$ may increase with the increase of both $f_1$ and $f_2$ even when another parameter does not change.

Thus, one can re-fit the Model 2 with latest cosmology observational data, supposing that $f_1>-0.34\times10^{-2}$ and $f_2>-6.56\times10^{-2}$, as is shown in Table I.  And the new best-fit parameters  is presented in Table \ref{table:comp1} and Table \ref{table:comp2}.

\begin{table}[h]
	\centering
	\begin{tabular}{c|c|c|c}
		\hline
		\hline
		\multirow{2}{*}{  Parameters} & \multicolumn{3}{c}{ Parametric Models  } \\
		\cline{2-4}
		& Model 2 &  Model 3&  Model 4   \\
		\hline
		\hline
		$\Omega_{m0}$
		& $0.255\pm 0.007$  	& $0.257\pm0.010$ &  $0.256\pm 0.009$ \\
		\hline
		$f_0$
		& $4.47\pm0.12 $  	& $4.45\pm0.10$ &  $4.47\pm 0.09$ \\
		$f_1(10^{-2})$
		& $-0.27\pm2.51$  	& $-$ &  $-0.36\pm0.77 $\\
		$f_2(10^{-2})$
		& $-0.08\pm7.07 $  	& $-0.66\pm2.10$ &  $-$ \\
		$f_{12}$
		& $-$ 	& $-$ &  $-$ \\
		$f_{14}$
		& $-$  	&$-$ &  $-$\\
		$f_3$
		& $0.77\pm 0.02$  	& $0.77\pm0.03$ &  $0.77\pm 0.03$ \\
		\hline
		$H_0$
		& $68.59\pm 1.21$ & $68.55\pm1.24$ & $68.55\pm 1.28$ \\
		\hline
		$\alpha$
		& $0.141\pm 0.006$ & $0.141 \pm 0.006$& $0.141\pm 0.006$ \\
		$\beta$
		& $3.102\pm 0.075$ & $3.103\pm0.075$ & $3.102\pm 0.074$\\
		$M_B$
		& $-19.10\pm 0.027$ & $-19.09\pm0.023$  &$-19.09\pm 0.026$ \\
		$\Delta M$
		& $-0.070\pm 0.023$ & $-0.069\pm0.024$ & $-0.070\pm 0.023$ \\
		\hline
		\hline
		$\chi^2_{min}/d.o.f$
		& $683.813/737$ & $683.815/738$  & $683.836/738$\\
		\hline
		\hline
	\end{tabular}
	\caption{\label{table:comp1} Best fitting parameters for the Model 2-4.}
\end{table}

\begin{table}[h]
	\centering
	\begin{tabular}{c|c|c|c|c}
		\hline
		\hline
		\multirow{2}{*}{  Parameters} & \multicolumn{4}{c}{ Parametric Models  } \\
		\cline{2-5}
		 &  Model 5 &  Model 6 &  Model 7 &  Model 8   \\
		\hline
		\hline
		$\Omega_{m0}$
		&$0.283\pm 0.010$&$0.264\pm 0.009$&$0.258\pm 0.002$&$0.221\pm 0.002$ \\
		\hline
		$f_0$
		&$-$&$-$&$-$&$-$ \\
		$f_1(10^{-2})$
		&$-$&$0.015\pm 0.229$&$-$&$0.015\pm 0.066$ \\
		$f_2(10^{-2})$
	&$-$&$0.032\pm0.061 $&$-$&$0.047\pm 0.021$ \\
		$f_{12}$
	&$1.69\pm 0.20$&$1.41\pm 0.23$&$-$&$-$ \\
		$f_{14}$
		&$-$&$-$&$1.98\pm 0.19$&$2.41\pm 0.28$ \\
		
		$f_3$
		&$0.85\pm 0.03$&$0.79\pm 0.02$&$0.77\pm 0.01$&$0.66\pm 0.06$ \\
		\hline
		$H_0$
		&$63.55\pm 0.63$&$67.11\pm 0.68$&$63.65\pm 0.12$&$65.28\pm0.42 $\\
		\hline
		$\alpha$
	 &$0.139\pm 0.006$&$0.140\pm 0.006$&$0.140\pm 0.006$&$0.141\pm 0.006$\\
		$\beta$
		&$3.096\pm 0.074$&$3.088\pm 0.075$&$3.080\pm 0.074$&$3.100\pm 0.074$ \\
		$M_B$
	 &$-19.19\pm 0.024$&$-19.21\pm 0.0178$&$-19.09\pm 0.041$&$-19.09\pm 0.034$\\
		$\Delta M$
		 &$-0.072\pm 0.029$&$-0.072\pm 0.023$&$-0.070\pm 0.002$&$-0.0.070\pm 0.023$\\
		\hline
		\hline
		$\chi^2_{min}/d.o.f$
	&$705.612/739$&$686.530/737$&$704.125/739$&$685.358/737$\\
		\hline
		\hline
	\end{tabular}
	\caption{\label{table:comp2} Best fitting parameters for the Model 5-8.}
\end{table}

\begin{table}[h]
	\begin{tabular}{ccccccc}
		\hline\hline
		\textbf{}           && \textbf{$\gamma-1$} &&Best Fit&&Within 1$\sigma$\\ \hline\hline
		\textbf{Model $2$}  &&    $(-1.7810\pm0.0892)\times 10^{-6} $ && $\checkmark$      &&  $-$     \\ 
		\textbf{Model $3$}  &&    $( -1.3128\pm 0.0480)\times 10^{-6}$  &&  $\checkmark$   && $- $      \\ 
		\textbf{Model $4$}  &&   $( -1.1759 \pm0.0663)\times 10^{-6}  $ &&  $\checkmark$    &&   $- $    \\ 
		\textbf{Model $5$}  &&   $ (-2.1524\pm0.4782)\times 10^{-6} $  &&    $\bigcirc$   &&$\checkmark$    \\ 
		\textbf{Model $6$}  &&    $ (-1.9752 \pm0.1300)\times 10^{-6} $   &&   $\checkmark$   && $-$   \\ 
		\textbf{Model $7$}  &&    $ ( -1.8679\pm0.1498)\times 10^{-6}$ &&  $\checkmark$  && $-$   \\ 
		\textbf{Model $8$}  &&   $( -1.9502\pm0.1101) \times 10^{-6} $&&  $\checkmark$   && $-$     \\ \hline\hline
	\end{tabular}
	\caption{The values of PPN parameters and their $1\sigma$ errors,  where we use the values from  Table VII in the calculation.\label{tab:riy}}
\end{table}

From Table \ref{table:comp1}, Table \ref{table:comp2} and \ref{tab:riy} one can see that  Model 2, 3, 4, which generalizes LCDM(Model 1) by adding intermediate power terms of $(1+z)$, are well agreed with cosmological background data and  the $1\sigma$ errors of $f_1$ and $f_2$ are also consistent with $f_1=0$ and $f_2=0$, when they go back to LCDM model. And they are also well favored with solar system tests. 

However, Model 5, 6, 7, 8, which modifies LCDM(Model 1) by taking the place of $f_0$ by adding $(1+z)^{-1/2}$ or $(1+z)^{-1/4}$ terms, are not so well agree with cosmological background data than Model 2, 3, 4 when solar system constraints are taken account in the data fitting. Comparing with Model 2, 3, 4, these models cannot go back to LCDM model when setting $f_1=f_2=0$ .

\section{Reconstruction of the $f(R)$ Actions from $f(z)$ parametric models}\label{sec:rec}

We have analysis the models individually to cosmology data and solar system data. So in this section, we may try to reconstruct the functions of $f(R)$ theories from the $f(z)$ parametric models above.  

According to Ref.\cite{Capozziello:2006jj} and Eqs.(\ref{gb1}), by defining $A=\frac{1-\gamma}{2\gamma-1}$, one can obtain that
\begin{eqnarray}
 f_{2R} =A f_R\,.
\end{eqnarray}
And the non-trivial solution to the differential equation reads
\begin{eqnarray}\label{eq:sa}
f(R)=\frac{1}{12} A R^3 \pm \frac{1}{2} \sqrt{A} R^2 +R-2\Lambda\,.
\end{eqnarray}
On the other hand, one can perform Taylor expansion about the Ricci scalar at $z=0$:
\begin{eqnarray}
f(R)=\sum_{n=0}^m g_n(R-R_0)^n\,,
\end{eqnarray}
where $g_n=\frac{1}{n!}f_{nR}\big|_{R=R_0}$ and $R_0=R(z=0)$. From Eqs.(\ref{eq:eom1})-(\ref{eq:eom2}), one can obtain that
\begin{eqnarray}
f(R_0)=6H_0^2(1-\Omega_{m0})+R_0\,.
\end{eqnarray}
Eqs.(\ref{eq:eom1})-(\ref{eq:eom2}) can be written as
\begin{eqnarray}\label{eq:third}
\frac{1}{2}f+(3\dot{H}-3H^2)f_R+3H\dot{R}f_{2R}&=&\rho_m\,,\\
\frac{1}{2}-(3H^2-\dot{H})f_R+(2H\dot{R}+\ddot{R})f_{2R}+\ddot{R}f_{3R}&=&0\,,
\end{eqnarray}
Eq.(\ref{eq:third}) can be rewriten as a third order differential equation of $f(R(z))$,
\begin{eqnarray}
A_3(z)f_{3z}+A_2(z)f_{2z}+A_1(z)f_z=-3H_0^2\Omega_{m0}(1+z)^3\,,
\end{eqnarray}
where $A_i(z)$ is functions consists $H(z)$ and its derivatives. Thus, according to Ref.\cite{Capozziello:2005ku,Feng:2008hk}, the effective gravatitional constant $G/f_R\big|_{R=R_0}$ must be equal to $G$ at $z=0$, so we get
\begin{eqnarray}
f_R\big|_{R=R_0}=f_zR_z^{-1}\big|_{z=0}\,.
\end{eqnarray}
and
\begin{eqnarray}
f_z\big|_{z=0}&=&R_z\big|_{z=0}\,,\\
f_{2z}\big|_{z=0}&=&R_{2z}\big|_{z=0}\,.
\end{eqnarray}

Thus, with the best fitting values in Table \ref{table:comp1} and Table \ref{table:comp2}, the $f(R)$ functions can be reconstructed in the similar form of Eq.(\ref{eq:sa}):
\begin{eqnarray}
f(R)_{\textbf{Model}}=R+b(R-R_0)^2+c(R-R_0)^3-2\Lambda^\prime\,,
\end{eqnarray}
where $b$ and $c$ are constants and $R_0$ is the present value of Ricci scalar which can be obatained by basic cosmological parameters in Eq.(\ref{eq:R0}). And the value of $\Lambda^\prime$ depends on other constant terms. Here we take Model 2 Table \ref{table:comp1} as an example:
\begin{eqnarray}
f(R)_{\textbf{Model 2}}=(R-R_0)-\frac{0.0016}{H_0^2}(R-R_0)^2-\frac{0.0069}{H_0^2}(R-R_0)^3-2.2950H_0^2\,.
\end{eqnarray}
where the  nonlinear terms could constitute a geometrical explanation for the expansion of the Universe.

\section{Conclusion and Discussion}\label{sec:conclusion}

In this paper, we have performed the solar system tests to the $f(z)$  parametric models, which are proposed to explain the accelerating expansion of the universe in Ref.\cite{Lazkoz:2018aqk}. After solving the equation for the Ricci scalar (\ref{eq:eqr}) numerically with $f(z)$  given by Model 2-8 in Eqs.(\ref{eq:m2})-(\ref{eq:m8}), we calculate the PPN parameters and compare them to recent data. We find that the $f(z)$ parametric models with constant term $f_0$ is much more favored by the solar system observations. And models without $f_0$ but with $(1+z)^{1/4}$ terms also satisfy the local constraints, while Model 5-6 with $(1+z)^{1/2}$ terms are a bit less favored than other models. According to the fitting values of the parameters in Model~2-4 in Ref.\cite{Lazkoz:2018aqk}, one could see that $f_1\sim1.0\times10^{-4}$ is much smaller than $f_3\sim0.94$, so Model~2-4 has a slightly difference to Model~1, i.e., the $\Lambda$CDM model. For example, in the future, the second term of Model 4 in Eq.(\ref{eq:m4}) becomes much more important than the third one, i,e.
\begin{equation}
1+z < \sqrt{\frac{f_1}{f_3}}-1 \approx 0.015\,.
\end{equation}
However, the constant term $f_0=4.43$ is already much more larger than the second one at the same time. So the $f_1$ term may not be important. And it is the same with other models.  After we reconstruct these models, we find that the difference between this group of models and general relativity is very small, and it will gradually show the difference when Ricci scalar is large enough. 

A general conclusion we can extract from this work is that $f(z)$ theories, as the authors of Ref.\cite{Lazkoz:2018aqk} stressed, may not be the definitive answer to explain why the Universe is accelerated expanding, but it provides a different and interesting perspective on how to relate the modified gravity with observations. And the parametric models, especially Model 2, 3, and 4, which can be considered as the simplest example of extended theories of LCDM(Model 1), succeed in addressing the phenomenology of solar system and Cosmology.

It should be noted that the reconstruction of the models are just approximation approaches. In other words, what we get in this paper is just some effective $f(R)$ functions to describe the behaviors of $f(z)$ parametric models. And of course, we can also reconstruct the $f(R)$ functions into other form or by higher order parameters such as the decelaration parameter\cite{Masoudi:2015ega,Carloni:2010ph}.  For example, as is well known, any function can be expanded by Pade approximation which can avoid the divergence problem in the higher order terms like Taylor expansion, which acquires further discussions. And the future data of BepiColombo Mission\cite{Serra:2018irk} will improve the precision of PPN parameter and may help us to test the theories of gravitation.

\acknowledgments
This work is supported by National Science Foundation of China grant Nos.~11105091 and~11047138, ``Chen Guan'' project supported by Shanghai Municipal Education Commission and Shanghai Education Development Foundation Grant No. 12CG51, and Shanghai Natural Science Foundation, China grant No.~10ZR1422000.

\end{document}